\documentstyle[epsf]{article}

\newcommand{\be}{\begin{equation}}
\newcommand{\ee}{\end{equation}}
\newcommand{\bea}{\begin{eqnarray}}
\newcommand{\beas}{\begin{eqnarray*}}
\newcommand{\eea}{\end{eqnarray}}
\newcommand{\eeas}{\end{eqnarray*}} 
\newcommand{\ba}{\begin{array}}
\newcommand{\ea}{\end{array}}
\newcommand{\bi}{\begin{itemize}}
\newcommand{\ei}{\end{itemize}}
\newcommand{\ben}{\begin{enumerate}}
\newcommand{\een}{\end{enumerate}}

\newcommand{\prl}{Phys. Rev. Lett.}
\newcommand{\prd}{Phys. Rev. {\bf D}}
\newcommand{\pl}{Phys. Lett.}
\begin{document}

\renewcommand{\baselinestretch}{1.5}
\large

%\twocolumn
%[\hsize\textwidth\columnwidth\hsize\csname @twocolumnfalse\endcsname]

\title{Non SUSY Unification in Left-Right Models.}
\author{Abdel P\'erez--Lorenzana$^1$, William A. Ponce$^{1,2}$
and Arnulfo Zepeda$^1$}

\date{\today}

\maketitle

\begin{center}
\normalsize 1-Departamento de F\'\i sica, 
\normalsize Centro de Investigaci\'on y de Estudios Avanzados del I.P.N.\\
\normalsize Apdo. Post. 14-740, 07000, M\'exico, D.F., M\'exico.\\
\normalsize 2-Departamento de F\'\i sica, Universidad de Antioquia 
\normalsize A.A. 1226, Medell\'\i n, Colombia.
\end{center}

\begin{abstract}
We explore in a model independent way the possibility of achieving the non
supersymmetric gauge coupling unification within left-right symmetric
models, with the minimal particle content at the left-right mass scale 
which could be as low as 1 TeV in a variety of models, and with a
unification scale M in the range $10^5$ GeV $< M< 10^{17.7}$ GeV.

\end{abstract}

Pacs: {11.10.Hi;12.10.-g;12.10.Kt}

\vskip2em

\section{Introduction}
It has been known for more than a decade\cite{amaldi} that if we let the
three gauge couplings $c_i\alpha_i^{-1}$  run through the ``desert" from
low to high energies, they do not merge together into a single point, 
where $\{c_1,c_2,c_3\}=\{\frac{3}{5},1,1\}$ are the normalization constants
of the Standard Model (SM) factors U(1)$_Y$, SU(2)$_L$ and SU(3)$_c$, 
respectively, embedded into SU(5)\cite{georgi}.
This odd result claims for new physics at intermediate energy scales as
for example:
\begin{enumerate}
\item The inclusion of the minimal supersymmetric (SUSY) partners of the
SM fields at an energy scale $M_{susy}\sim 1$ TeV, related to an 
unification scale $M\sim 10^{16}$ GeV\cite{boer}.
\item The inclusion of a minimal Left Right Symmetric Model (LRSM) at a
mass scale $M_R\sim 10^{11}$ GeV, related to an unification scale $M\sim
10^{15}$ GeV\cite{shaban} in an SO(10) Grand Unified Theory (GUT)
\cite{geor10}.
\item The inclusion of the SUSY partners of the minimal LRSM 
at an energy scale $M_{susy}\sim M_R\sim 1$ TeV,
related to a unification scale $M\sim 10^{16}$ GeV\cite{rizzo}. Etc..
\end{enumerate}
The alternative approach, namely, to normalize the gauge couplings
$c_i\alpha_i^{-1}$ to non-orthodox $c_i,\; (i=1,2,3)$ values  was
presented by these authors, in Ref. \cite{abdel1} for non-SUSY models, and
in Ref. \cite{abdel2} for the SUSY ones, for possible GUT models which can
descend in one single step to $SU(3)_c\otimes SU(2)_L\otimes
U(1)_Y\equiv G_{SM}$.

In this paper we present a systematic analysis of all the possible GUT models
which descend in two steps to $G_{SM}$, with the 
LRSM as the intermediate step, paying special attention to those models
with low $M_R$ scale. The paper is organized in the following way:
In section II we present the renormalization group equation formalism for
the LRSM; in section III we carry our model independent analysis, and in
section IV we present our results and  conclusions. A technical appendix at the
end gives the $c_i,\; i=1,2,3$ values for most of the GUT models in the
literature.

%%%%%%%%%%%%%%%%%%%%%%%%%%%%%%%%%%%%%%%%%%%%%
\section{The Renormalization Group Equations}
%%%%%%%%%%%%%%%%%%%%%%%%%%%%%%%%%%%%%%%%%%%%%

In a field theory, the couplings are defined as effective
values, which are  energy scale dependent according to the
renormalization group equations. 
In the  modified minimal substration scheme ~\cite{ms},
which we adopt in what follows, the one-loop renormalization group
equations are
\be
\mu{d\alpha_i\over d\mu} \simeq -b_i \alpha_i^2, \label{rge}
\ee
where $\mu$ is the energy at which the coupling constants $\alpha_i =
g_i^2/4\pi, \, (i=1,2,3)$ are evaluated, with $g_1$, $g_2$, and $g_3$ the
gauge couplings of the 
SM factors $U(1)_Y$, $SU(2)_L$ and $SU(3)_c$ respectively. The
constants $b_i$ are completely determinated by the  particle content in the
model by
\[
4\pi b_i = {11\over 3} C_i(vectors) - {2\over 3}C_i(fermions)
-{1\over 3}C_i(scalars), 
\]
being $C_i(\cdots)$ the index of the representation to which the $(\cdots)$
particles are assigned, and where we are considering Weyl fermion and
complex scalar fields~\cite{bs}. The boundary conditions for these equations 
are determined by the relationships
\be
\alpha^{-1}_{em} = \alpha_1^{-1}+ \alpha_2^{-1},  \quad\mbox{and}\quad
\tan^2\theta_W = {\alpha_1\over\alpha_2}, \label{rel1}
\ee
which at the electroweak scale imply
\be
\alpha_1^{-1}(m_Z)=\frac{1-{\rm sin}^2\theta_W(m_Z)}{\alpha_{em}(m_Z)},
\quad\mbox{and}\quad
\alpha_2^{-1}(m_Z)=
\frac{{\rm sin}^2\theta_W(m_Z)}{\alpha_{em}(m_Z)}.\label{reel1}
\ee

Combining those expressions with the experimental values 
\bea
 \alpha^{-1}_{em}(m_Z) &=& 127.90 \pm 0.09~\cite{pdg,ewk},\nonumber\\  
\sin^2\theta_W (m_Z)&=& 0.2312 \pm 0.00017~\cite{pdg,ewk} \quad \mbox{and}
\label{data}\\
\alpha_3(m_Z)&=& \alpha_s = 0.1191\pm 0.0018~\cite{pdg},\nonumber
\eea

we get:
\bea
\alpha^{-1}_1(m_Z) &=& 98.330\pm 0.091, \nonumber\\
\alpha^{-1}_2(m_Z) &=& 29.571\pm 0.043, \quad\mbox{and} \label{alphas}\\
\alpha^{-1}_3(m_Z) &=&  8.396\pm 0.127. \nonumber
\eea

The unification of the three SM gauge couplings is properly achieved
if they meet together into a common value $\alpha=g^2/4\pi$ at a
certain energy scale $M$, where $g$ is the gauge coupling constant of
the unifying group $G$. However, since $G\supset G_{SM}$,  
the normalization of the  generators corresponding  to the subgroups
$U(1)_Y$, $SU(2)_L$ and $SU(3)_c$ is in general different for
each particular group $G$, and therefore the SM  coupling constants
$\alpha_i$ differ at the  
unification scale from $\alpha$ by numerical factors $c_i\, (\alpha_i
=c_i\alpha )$. In $SU(5)$ these factors are\cite{georgi} 
$\{c_1,c_2,c_3\} = \{{3\over 5},1,1\}$ (we call them the canonical
values), which are the same for 
$SO(10)$~\cite{geor10}, $E_6$~\cite{e6}, $[SU(3)]^3\times Z_3$\cite{su33},  
$SO(18)$\cite{so18}, $E_8$\cite{e8}, $SU(15)$\cite{su15}, 
$SU(16)$\cite{su16}, and $SU(8)\otimes SU(8)$\cite{2su8}, but they are 
different for other groups such as $SU(5)\otimes SU(5)$~\cite{2su5}, 
$[SU(6)]^3\times Z_3$\cite{3su6}, the Pati-Salam models\cite{patis}, 
etc. (see Table I in the appendix).

The constants $c_i$ can also be seen as a consequence of the affine 
levels (or Kac-Moody levels) at which the gauge factor $G_i$ is 
realized in the effective four dimensional string\cite{kacmody}, even 
if there is not an unification gauge group at all;
but if it does, they are related to the fermion content of the 
irreducible representations of $G$. As a matter of fact, if
$\alpha_i$ is the coupling constant of $G_i$, a simple group embedded
into $G$, then 
\be 
c_i\equiv {\alpha_i\over \alpha}= {Tr\ T^2\over Tr\ T^2_i}, \label{ci}
\ee
where $T$ is a generator of the subgroup $G_i$ properly normalized over a  
representation $R$ of $G$, and $T_i$ is the same generator but normalized
over the representation of $G_i$ embedded into $R$ (the traces run over
complete representations). In this way for example, if just one standard
doublet of $SU(2)_L$ is contained in the fundamental representation  of
$G$ (plus any number of $SU(2)_L$ singlets), then $c_2 = 1$ (as in
$SU(5)$); but this is not the general case. In this way we proof that 
for $i=2,3$, $c_i^{-1}=1,2,3,...n$ an integer number. 
The constants $c_i$ are thus pure rational numbers satisfying 
$c_1>0$ and $0<c_{2(3)}\leq 1$. They are 
fixed once we  fix the unifying gauge structure. 
According to the table I in the appendix and in order to 
simplify matters,  we are going to use for $c_2$ only the 
values $1$ and ${1\over 3}$, 
and for $c_3$ the values 1 and ${1\over 2}$.

From Eqs. (\ref{rel1}) and (\ref{ci}) it follows that at the unification scale
the value of $\sin^2\theta_W$ is given by
\be
\sin^2\theta_W \equiv {\alpha_{em}\over\alpha_2} = {c_1\over c_1 + c_2}.
\label{sin2}
\ee
Obviously, Eq. (\ref{sin2}) is equivalent to that given in terms of the
traces of the generators of $SU(2)_L$ and the electric charge for simple 
groups (see Ref.~\cite{georgi}).
In order to connect this value at the scale $M$ with the corresponding value at
the scale $m_Z$ the renormalization group equations (\ref{rge}) must be
solved. 

Our approach is now the following: we assume there are only three
relevant mass scales $m_Z,\, M_R$, and $M$ such that $m_Z<M_R<M$, where
$m_Z\sim 10^2$GeV is the electroweak mass scale, $M_R$ is the mass
scale where the LRSM (with and without discrete left-right (LR) symmetry) 
manifests itself, and $M$ is the GUT scale. Then, the equations
(\ref{rge})  must be solved,
first for the energy range $m_Z< \mu < M_R$, and then for the range 
$M_R < \mu < M$, properly using at each stage the decoupling theorem
\cite{appel}. 

Now for the energy interval $m_Z< \mu <M_R$, the one loop solutions to the
equations (\ref{rge}) are:
\be 
\alpha_i^{-1}(m_Z) = \alpha_i^{-1}(M_R)- 
b_i(H)\, \ln\left({M_R\over m_Z} \right), \label{solrge}
\ee
where the beta functions $b_i \, (i=1,2,3)$ are \cite{bs} 
\be 
2 \pi \left( \ba{c} b_1 \\[1ex] b_2 \\[1ex] b_3 \ea \right) = 
 \left( \ba{c} 0 \\[1ex] 22\over 3 \\[1ex] 11 \ea \right) - 
 \left( \ba{c} 20\over 9 \\[1ex] 4\over 3 \\[1ex] 4 \over 3 \ea \right) F - 
 \left( \ba{c}  1\over 6 \\[1ex]   1\over 6 \\[1ex] 0 \ea \right) H,\label{b}
\ee
with $F=3$ the number of families and $H$ the number of low energy 
Higgs field doublets. Notice by the way that we
are not including in the former equation the
normalization factor $3\over 5$ into $b_1$ coming from
$SU(5)$, and wrongly included in some general discussions.
$H=1$ in the SM; nevertheless, a general model can have  
more than one low energy Higgs field, and in principle $H$ may be taken
as a free parameter ($H=2$ in the minimal supersymmetric model).

For the interval $M_R<\mu<M$, the evolution of the gauge couplings is
dictated by the beta functions of the LRSM whose gauge group is
\cite{leftright} 
$G_{LR}\equiv SU(3)_c\otimes SU(2)_L\otimes SU(2)_R\otimes U(1)_{B-L}$, 
with the matter fields transforming as $\Psi_L
= (3,2,1,1/6)\oplus (\bar 3,1,2,-1/6)\oplus (1,2,1,-1/2)\oplus
(1,1,2,1/2)$ for each generation, where the numbers between brackets 
label ($SU(3)_c, SU(2)_L$, $SU(2)_R$, $U(1)_{B-L}$) representations.

The LRSM is broken down spontaneously by the Higgs sector, which in
general contains $N_B$ bidoublet Higgs fields $\varphi(1,2,2,0)$, $N_{TL}$
triplets in the representation $\Delta_L(1,3,1,1)$, $N_{TR}$
triplets in the representation $\Delta_R(1,1,3,-1)$, $N_{DL}$ doublets in
the representation $\phi_L(1,2,1,-1/2)$, and $N_{DR}$ doublets in the
representation $\phi_R(1,1,2,1/2)$. 
In the so called minimal LRSM\cite{leftright}, $ N_{TR}=N_B=1$ and
$N_{DL}=N_{DR}=N_{TL}=0$, but in general $N_{TL},\, N_{TR},\, N_{DL},\,
N_{DR}$ and $N_B$ should be taken as free parameters to be fixed by the
specific model.  

In a general context, the vacuum expectation values that may be used to 
break the symmetry are 
$\langle\Delta_R^0\rangle\sim\langle\phi^0_R\rangle\sim
M_R\; (\Delta_R^0$ represent the electromagnetic neutral direction in
$\Delta_R$, etc.), 
$\langle \varphi^0\rangle\sim\langle \phi^0_L\rangle\sim m_Z$, and
$\langle \Delta^0_L\rangle=0$. It then follows that $H=2N_B + N_{DL}$. 

The discrete LR symmetry implies invariance under the exchange
$L\leftrightarrow R$ in the model (this is the so called D parity) with the
consequence that $g_{2L}=g_{2R}$ for the energy interval $M_R<\mu<M$. This
symmetry is respected by the gauge and the fermion content of any LRSM,
but it is broken by the scalar sector as it is shown anon.

Indeed, the Higgs field scalars can drastically alter the solution to the
renormalization group equations, and in order to make any definite
statement about the mass scales in a particular model, we must know which
components of the Higgs representations have masses of order $m_Z,\; M_R$
and $M$. However, to know the masses of the scalars is equivalent to the
hopeless task of knowing the values of all the coupling constants
appearing in the scalar potential (with radiative corrections included).
So, in order to guess what the real effect of the scalars is, the so
called extended survival hypothesis was introduced in Ref.~\cite{esh}.
Basically the hypothesis consists in assuming that only the components of
the Higgs representations which are required for the breaking of a
particular symmetry are the only ones which are not superheavy. In other
words: ``scalar Higgs fields acquire the maximum mass compatible with the
pattern of symmetry breaking" (for a more detailed explanation and
application to SO(10), see Ref.~\cite{esh}).

The one loop solutions to Eqs. (\ref{rge}) for the energy
interval $M_R<\mu<M$ are:
\be
\alpha_i^{-1}(M_R) = {1\over c_i}\alpha^{-1} -
 b'_i(N_B,N_{TL},N_{TR},N_{DL},N_{DR})\, \ln\left({M\over M_R}\right),
\label{at}
\ee
where $i=BL,2L,2R,3$. The beta functions $b'_i$ are now:
$b'_3=b_3=7/2\pi$ (with the assumption that no low energy colored scalars
exist (as demanded by the extended survival hypothesis); if they do, they
may cause a too fast proton
decay, and spoil the asymptotic freedom for $SU(3)_c$); and $b'_{2R}, \,
b'_{2L}$ and $b'_{BL}$ given by: 
 \bea
 2 \pi \left( \ba{c} b'_{BL}\\[1ex] b'_{2L} \\[1ex] b'_{2R} \ea \right) &=& 
 \left( \ba{c} 0 \\[1ex] 22\over 3 \\[1ex] 22\over 3 \ea \right) - 
 \left( \ba{c} 8\over 9 \\[1ex] 4\over 3  \\[1ex] 4\over 3 \ea \right) F - 
\left( \ba{c}  0 \\[1ex] N_B\over 3 \\[1ex]   N_B\over 3 \ea \right) -
\left( \ba{c}  N_{TL}+N_{TR}\\[1ex]  \frac{2N_{TL}}{3}\\[1ex]
\frac{2N_{TR}}{3}\ea \right)-\left( \ba{c}
\frac{N_{DL}+N_{DR}}{6}\\[1ex] \frac{N_{DL}}{6} \\[1ex] \frac{N_{DR}}{6} \ea
\right).\label{b'}
\eea
From Eqs. (\ref{at}) and (\ref{b'}) we get $g_{2L}=g_{2R}$ if
$N_{TL}=N_{TR}$ and $N_{DL}=N_{DR}$. But if  $N_{TR}\neq N_{TL}=0$ as
demanded by the extended survival hypothesis, then one could only have
exact left-right symmetry at the
GUT scale.

The hypercharge $Y$ of the SM is given by 
\be
Y = T_{3R} + Y_{B-L},
\ee
which implys the relation 
$ \alpha_1^{-1}(M_R) = \alpha_{2R}^{-1}(M_R) + \alpha_{BL}^{-1}(M_R)$. 
Then the beta function for $U(1)_Y$ for the energy interval $M_R<\mu<M$
may be written as $b_1' = b'_{2R} + b'_{BL}$ with $c_1^{-1} = c_{2R}^{-1}
+c_{BL}^{-1}$, and $c_{2R}=c_{2L} = c_2$, ($c^{-1}_{BL}={2\over 3}$ for the
minimal fermion field content of the LRSM). These relations together with
Eqs. (\ref{solrge}) and (\ref{at}) allow us to write:
 \bea
\alpha_1^{-1}(m_Z)&=& {1\over c_1}\alpha^{-1} +
 \left(\frac{40+H}{12\pi}\right)\,\ln\left({M\over m_Z}\right)- 
\frac{1}{6\pi}(22-3N_{TL}-5N_{TR}-N_{DR})\,
\ln\left(M\over M_R\right) \nonumber \\
\alpha_2^{-1}(m_Z)&=& {1\over c_2}\alpha^{-1} -
\frac{1}{12\pi} \left[(20-H)\, \ln\left({M\over m_Z}\right) -
4N_{TL}\,\ln \left({M\over M_R}\right)\right] \label{3eqs} \\
\alpha_3^{-1}(m_Z)&=& {1\over c_3}\alpha^{-1} 
-\frac{7}{2\pi}\, \ln\left({M\over m_Z}\right),\nonumber 
\eea
which is a system of 3 equations with 
3 unknowns: $\alpha,\; M_R$ and $M$ ($m_Z=91.187\pm 0.007$  
GeV\cite{pdg} and $\alpha_i^{-1}(m_Z)$ as in Eqs. (\ref{alphas}) are
taken as imputs). $c_i (i=1,2,3)$, $N_B,N_{TL}, N_{TR}, 
N_{DL}$, and $N_{DR}$ $(H=2N_B+N_{DL})$ are model dependent parameters. 
Evidently, there is always solution to the system of equations in
(\ref{3eqs}), but the consistency of the unification scheme demands that 
$m_Z<M_R<M\leq10^{19}$ GeV$\sim M_P$ (the Planck Mass). When we solve Eqs.
(\ref{3eqs}) for the minimal LRSM $(N_{TR}=N_B=1,\,
N_{TL}=N_{DL}=N_{DR}=0)$ for the canonical values
$(\{c_1,c_2,c_3\}=\{{3\over 5},1,1\})$ we get $M=2.5\; 10^{16}$ GeV,
$M_R=2.7\; 10^9$ GeV and $\alpha^{-1}=45.45$. 

Notice that if $N_{TL}=0$ (as demanded by the extended survival
hypothesis), the last two equations
in (\ref{3eqs}) are independent of $M_R$, and they are enough to fix the
GUT scale $M$ (and $\alpha$ of course). 
If we solve them for $c_2=1,\, c_3={1\over 2}$ (one family models with
chiral color\cite{2su3}, as for example
$SU(5)\otimes SU(5)$\cite{2su5}, $SO(10)\otimes SO(10)$ \cite{2so10}),  
we get for $H<22$ the unphysical solution $M>>M_P$. A further analysis shows
that for $22<H<30$ we get $M_P>M>10^{16}$ GeV which in turn implies $M_R<m_Z$
which is also unphysical. To get $M_R>1$ TeV requires for those models $H>40$
which gives $M<10^{12}$ GeV in serious conflict with proton decay which is
always present in those models. So the two step breaking pattern  
$SU(5)\otimes SU(5)\longrightarrow G_{LR}\longrightarrow G_{SM}$ is not
allowed (the one step $SU(5)\otimes SU(5)\longrightarrow G_{SM}$ is also
forbidden\cite{abdel1,abdel2}). This conclusion is valid even for the case
$g_{2L}\neq g_{2R}$ at the GUT scale, a variant of the model introduced 
in the second paper of Ref. \cite{2su5}. Similar conclusions follow for  
$SO(10)\otimes SO(10)$~\cite{2so10}. To use $N_{TL}\neq 0$ makes things even
worse.

When we solve Eqs.(\ref{3eqs}) for  $c_2={1\over 3},\,
c_3=1$ (models with three families and vector-like color as for example
$[SU(6)]^3\times Z_3$ \cite{3su6})  we get $M\simeq 5m_Z$, an unacceptable 
solution. So the two step breaking pattern 
$[SU(6)]^3\times Z_3\longrightarrow G_{LR}\longrightarrow G_{SM}$ is not
allowed either (the one step breaking pattern 
$[SU(6)]^3\times Z_3 \longrightarrow G_{SM}$  
is also forbidden for this group\cite{abdel1,abdel2}). 

So our analysis makes sense only for two cases: $\{c_2,c_3\}=\{1,1\}$
(one family models with vector like-color), and $\{c_2,c_3\}=\{{1\over
3},{1\over 2}\}$ (models with 3 families and chiral color). In what follows 
we are going to refer only to these situations.

Before moving to a general analysis, let us see for example what happens for
$SO(10)\longrightarrow G_{LR}\longrightarrow G_{SM}$.  As mentioned above,
$\{c_1,c_2,c_3\}=\{{3\over 5},1,1\}$, and there are not exotic fermions in the
spinorial 16 representation used for the matter fields, but the scalar
content is not
quite uniquely defined, and there are as  many versions of the model as you
wish. A couple of examples are:\\
1- In Ref.\cite{shaban} the following symmetry breaking pattern is implemented:

\[ SO(10)\stackrel{\phi_{(210)}}{\longrightarrow} G_{LR}
\stackrel{\phi_{(126)}}{\longrightarrow} G_{SM}
\stackrel{2\phi_{(10)}}{\longrightarrow} SU(3)_c\otimes U(1)_{EM}.\]

\noindent
$\phi_{210}$ gets mass at the GUT scale and it does not contribute to the
renormalization group equations. 
For $\phi_{126},\; \Delta_R=\Delta_L=1$, but only $\langle\Delta_R\rangle\neq
0$.  For the final breaking only one $\phi_{(10)}$ is needed, but at least two
must be used  in order to achieve proper isospin breaking. Then $N_{TR}=1,\;
N_B=2,\; N_{TL}=N_{DL}=N_{DR}=0$. We get $M=2.0 \; 10^{15}$ GeV, $M_R=1.6 \;
10^{11}$ GeV and $\alpha^{-1}=42.6$. \\
2- A more recent version of (SUSY) S0(10) implements the breaking with the
following scalar content\cite{barr}:
\[ SO(10)\stackrel{\phi_{(45)}}{\longrightarrow} G_{LR}
\stackrel{\phi_{(16+cc)}}{\longrightarrow} G_{SM}
\stackrel{2\phi_{(10+16+cc)}}{\longrightarrow} SU(3)_c\otimes U(1)_{EM}.\]
\noindent
With the extended survival hypothesis in mind we have $N_{TL}=N_{TR}=0,\;
N_B=N_{DL}=N_{DR}=2$. We get 
$M=2.2 \;10^{14}$ GeV, $M_R=9 \; 10^{12}$ GeV, and $\alpha^{-1}=40.16$. 
In both examples the D parity is broken below the GUT scale.  

Since the scalar sector is the most obscure part of any gauge theory, it is
clear that, $N_i\; (i=B,TL,TR,DL$ and $DR)$ can be taken as free parameters,
resulting in all sort of models for all sort of tastes. Since the Higgs
field scalars can drastically change the GUT scales, we can not state with
confidence neat values for $M$ and $M_R$. We elaborate on this in the next
section. 

Before proceeding to our model independent analysis let us mention that we
are going to consider the possibility of adding arbitrary large numbers of
scalars Higgs fields in order to get unification. In many cases this
may result in the coupling constants becoming so large as to make the
theory non-perturbative before unification is achieved. Even though the
extended survival hypothesis\cite{esh} greatly diminishes the effect of the
Higgs scalar fields, we will pay special attention to our parameter space
region in the analysis, in order not to run into non-perturbative regimes
of the coupling constants. As a mater of fact, the assumption that no low
energy colored scalars exist is all what is needed for the cases 
considered ahead.

%%%%%%%%%%%%%%%%%%%%%%%%%%%%%%%%%%%%%%%%%%%%%
\section{Model Independent Analysis}
%%%%%%%%%%%%%%%%%%%%%%%%%%%%%%%%%%%%%%%%%%%%%

In this section we are going to study two different situations.  First we are
going to reduce the freedom we have in our parameter space by imposing the
extended survival hypothesis. Second, we reduce the freedom by restoring
the D parity to the LRSM.

%%%%%%%%%%%%%%%%%%%%%%%%%%%%%%%%%%%%%%%%%%%%%
\subsection{Solutions to the equations with extended survival hypothesis}
%%%%%%%%%%%%%%%%%%%%%%%%%%%%%%%%%%%%%%%%%%%%%

If we impose the extended survival hypothesis as a constraint in the
solutions to the renormalization group equations for the LRSM, 
we must set $N_{TL} = 0$. Then Eqs. (\ref{3eqs}) get reduced to a system of 3
equations with 3 unknowns, and the  following set of parameters: $c_i$ 
$(i = 1; 2; 3)$; $H$, and $N'_T= 5N_{TR} + N_{DR}$. 
The solution of Eqs. (\ref{3eqs}) for $M$, $M_R$ and
$\alpha$ as functions of these parameters is:
 \bea
\alpha^{-1} &=& \frac{42t_{32}-(20-H)t_{23}}{D} \label{a} \\[1ex]
\ln \left({M\over m_z}\right) &=&
\frac{12\pi}{D}\left[ c_2\alpha^{-1}_2(m_Z)-c_3\alpha^{-1}_3(m_Z)\right]
\label{xx}
\eea
 and
\be
\ln\left({M\over M_R}\right)=\frac{6\pi N}{c_1(22-N'_T)D}  \label{yy}
\ee
where  $N=[(20-H)(t_{21}-t_{23})+(40+H)(t_{12}-t_{13})+42(t_{32}-t_{31})]$,  $D
= 42c_3-(20-H)c_2$ and $t_{ij}=t_{ij}(m_Z)=c_ic_j\alpha^{-1}_j(m_Z)$. From Eq.
(\ref{yy}) it can be  seen that either $H<7$ and $N'_{T}<22$ ($N_{TR}<5$), 
or $H>7$ and $N'_T >22$, in order to have $M_R\leq M$.

From Eqs. (\ref{xx}) and (\ref{yy}) 
we plot in Figure 1 the allowed region for $H$ and $N'_T$ that give
unification, for the canonical values of $c_i$; and in Figure 2  we plot
$c_1$ Versus $N'_T$
for $H = 2$ and $\{c_2; c_3\} = \{{1\over 3} ; {1\over 2}\}$. 

To analyze the implications of each one of the figures we must have in mind 
the following constraints:
\begin{enumerate}
\item $M\leq M_P\sim 10^{19}$ GeV, the Planck scale (actually 
$M\leq M_{max}\sim 10^{17.7}$ GeV, obtained when there is not  contribution
from the scalar sector).
\item $M>10^5$ GeV in order to suppress unwanted flavor changing neutral
currents~\cite{pdg,akagi}.
\item $M>10^{16}$ if the proton is allowed to decay in the particular GUT
model.
\item  $8\, m_Z\leq M_R\leq M$. The lower limit is taken from the particle data
book\cite{pdg}, the upper limit is imposed by consistency of the
renormalization group equations.
\end{enumerate}

\subsubsection{Analysis of Figure 1.}
The allowed region lies inside the lines $M_R=8\, m_Z$ and $M_R=M$, but if the
proton does decay in the model under consideration then the allowed region
lies in the lower left corner between the lines $M=10^{16}$ GeV, $M_R=8\, m_Z$,
$H=0$ and $N'_T=0$. 

For GUT models with unstable proton (which are most of the models for the
groups in the canonical entry in Table 1 in the appendix), $M_R\sim 1$ TeV is
obtained for $H=2$ and $N'_{T}=13$ ($N_{TR}=2$ and $N_{DR}=3$), which in
turn implies $M\sim 2.59\; 10^{16}$ GeV.

For models in the canonical entry with a stable proton (as for 
example $[SU(3)]^3\times Z_3$\cite{su33} and $SU(8)\otimes SU(8)$\cite{2su8}) 
the allowed region is wider and divided in two regions. One for $H< 7$;
$N'_T< 22$ and the other for $H>7$ and $N_T>22$. There are plenty of
examples of models with $M_R\sim 1$ TeV for those situations. 

\subsubsection{Analysis of Figure 2.}
The entire plane in figure 2 is related to the GUT scale $M\sim 10^8$ GeV
(fixed just by the values of $H,\; c_2$ and $c_3$). The allowed region lies
between the lines $M_R=M$ and $M_R=8\, m_Z$. From the figure we see that a
value of $c_1={3\over 19}$ crosses the $M_R=8 \, m_Z$ line at $N'_T=31\;
(N_{TR}=6,\; N_{DR}=1)$, which means that the model $[SU(6)]^4\times
Z_4$\cite{4su6} can
have the following chain of spontaneous descent 

\[ [SU(6)]^4\times Z_4\stackrel{M}{\longrightarrow} G_{LR} 
\stackrel{M_R}{\longrightarrow} G_{SM} 
\stackrel{m_Z}{\longrightarrow} SU(3)_c\otimes U(1)_{EM}, \]

\noindent
with $M\sim 10^8$ GeV and $M_R\sim 9\, m_Z$, as long as an irreducible
representation of the GUT
group with 6 right handed triplets is used to break $G_{LR}$ down to the SM
gauge group and then a representation of the GUT group with only two
$SU(2)_L$ Higgs field doublets is used in the last breaking step. 

A further look into the equations for this group shows that for $N'_T=0$
and $H=2,3$ we get $M_R=M\sim 10^8$ GeV, meaning that a single step
spontaneous descent is possible for this model with a very economical set of
Higgs field scalars. But this result has been already published in
Ref.\cite{abdel1}. Here we just confirm the published result. 

%%%%%%%%%%%%%%%%%%%%%%%%%%%%%%%%%%%%%%%%%%%%%
\subsection{Solutions to the equations with D parity}
%%%%%%%%%%%%%%%%%%%%%%%%%%%%%%%%%%%%%%%%%%%%%

In order to restore the D parity in the renormalization group equations  
for the energy interval $M_R <\mu < M$ we must have
$N_{TL} = N_{TR}\equiv  N_T$ and $N_{DL} = N_{DR} = N_D$. Again we solve
Eqs. (\ref{3eqs}) as a function of $c_i$, $H$, $N_T$ and $N_D$. 
Using the equations we get, we plot in Figure 3 the allowed region for $H$
and $N_T$ that gives unification for the canonical 
values of $c_i$, and in Figure 4 we plot $c_1$ Versus $N_T$ for $H = 2$, 
$N_D = 0$ and $\{c_2; c_3\} = \{1; 1\}$. 

\subsubsection{Analysis of Figure 3}
For models with unstable proton the allowed tiny region lies in the lower
left corner, between the lines $N_T=0, \; H=0$ and $M=10^{16}$ GeV. From the
figure we get $M_R>10^9$ GeV, $N_T\leq 1$ and $H\leq 2$. 

For models with an stable proton the allowed region is larger, with
boundaries given by the lines $M_R=M$ and $M=10^5$ GeV which excludes the
possibility $M_R\sim$ a few TeV, unless $N_T>50$ which is very unlikely in
realistic models. 

\subsubsection{Analysis of Figure 4}
The allowed region of parameters lies inside the lines $N_T=0,\; M=M_R$ and
$M=10^{16}$ GeV for models with unstable proton, and inside the lines
$N_T=0,\; M=M_R$ and $M_R=8\, m_Z$ for models with an stable proton. As can be
seen, the canonical value $c_1={3\over 5}$ lies inside both regions, but
far from $M_R\sim 1$ TeV. 

In general, large values for $N_T$ are required $(H<8)$ in LRSM with D
parity, in order not to have unduly large values for $M_R$. 

%%%%%%%%%%%%%%%%%%%%%%%%%%%%%%%%%%%%%%%%%%%%%
\section{Conclusions}
%%%%%%%%%%%%%%%%%%%%%%%%%%%%%%%%%%%%%%%%%%%%%

To conclude let us emphasize that it is possible to unify the SM group using
the LRSM as an intermediate stage for a variety of models, with 1 TeV $\leq
M_R\leq 10^{15}$ GeV. From our  study, three family models with vector
like color
are excluded (as $[SU(6)]^3\times Z_3$), and one family models with chiral
color are also excluded (as  $SU(5)\otimes SU(5)$~\cite{2su5}, and
$SO(10)\otimes SO(10)$~\cite{2so10}).

We point out that in our analysis we  have neglected threshold  effects which
depend on the particular structure of each model, and also we do not include
second order corrections to the renormalization group equations which are
typically of the order of the
threshold effects. In others aspects it is completely general. Within this
limitations we may  conclude that it is indeed possible to achieve the
unification of the coupling constants of the SM  in a general class of non
supersymmetric models  which have the minimal LRSM as an intermediate step, 
with an $M_R$ scale as low as 1 TeV.  We are aware that this class of models
may  suffer of hierarchy problems.

From our analysis we may extract the following morals:\\
1- Higgs scalars play a crucial role in the solution to the
renormalization group equations.\\
2- It is simple to construct realistic non SUSY - GUT models with an
intermediate Left-Right symmetry at a mass scale $M_R\sim 1$ TeV (just read
them from the figures).\\
3- LRSM with D parity are quite different to those without D parity.\\
4- For low $M_R$, models with D parity are less realistic that models
without D parity, in the sense that they make use of a very large amount
of Higgs scalars.\\
5- It is impossible to sustain the D parity when the extended survival
hypothesis is imposed. 

%%%%%%%%%%%%%%%%%%%%%%%%%%%%%%%%%%%%%%%%%
\section*{Acknowledgments.}
%%%%%%%%%%%%%%%%%%%%%%%%%%%%%%%%%%%%%%%%%

We acknowledge R.N.Mohapatra for discussions
and  comments.  This work was partially supported by CONACyT, M\'exico.

%%%%%%%%%%%%%%%%%%%%%%%%%%%%%%%%%%%%%%%%%
\section*{Appendix.}
%%%%%%%%%%%%%%%%%%%%%%%%%%%%%%%%%%%%%%%%%

In this appendix we give the $c_i$ $i = 1; 2; 3$ values  for most of the GUT
groups in the literature. They are presented in table I.  The ``Canonical"
entry refers to the following  groups: $SU(5)$~\cite{georgi}
$SO(10)$~\cite{geor10}, $E_6$~\cite{e6}, $[SU(3)]^3 \times
Z_3$~\cite{su33}, $SU(15)$~\cite{su15}, $SU(16)$~\cite{su16}, $SU(8) \times
SU(8)$~\cite{2su8}, 
$E_8$~\cite{e8}, and $SO(18)$~\cite{so18}. Also, in the Canonical  entry we
have normalized the $c_i$ values to the $SU(5)$ numbers; for example, the
actual values for $SU(16)$ are: $\{c^{-1}_1; c^{-1}_2; c^{-1}_3\} = \{{20/ 3};
4; 4\} = 4\{5/3; 1; 1\}$. This normalization makes sense because 
physical quantities such as sin$^2\theta_W $, $M_R$ and $M$ depend only
on ratios of two $c_i$ values  (see Eqs. (\ref{sin2}), (\ref{xx}), and
(\ref{yy})).

Most of the groups in the first entry have the canonical values for $c_i$ due
to the fact that they contain $SU(5)$ via regular embeddings  (see the table 58
in Ref.\cite{slansky}), which do not change the rank of the corresponding
group. For others as for example $SU(16)$ it is just an accident.

$c^{-1}_3$ can take only the values $1,2,3,4$ for one family groups,  or higher
integer values for family groups. $c^{-1}_3 = 1$ when it is $SU(3)c$ which
is embedded in the GUT group $G$; $c^{-1}_3 = 2$ when it is the chiral
color~\cite{2su3} $SU(3)_{cL}\times SU(3)_{cR}$ which is embedded in $G$, etc.
For example $c^{-1}_3 = 4$ in $SU(16)$ due to the fact that the color group in
the GUT group is $SU(3)_{cuR}\times SU(3)_{cdR}\times SU(3)_{cuL}\times
SU(3)_{cdL}$.

For family groups $c^{-1}_2$ take the values $1, 2,\dots F$ for $1, 2, \dots
F$ families.  Indeed, the $c_i$ values for the $F$ family
Pati-Salam models~\cite{elias} $[SU(2F)]^4\times Z_4$ are $\{c^{-1}_1;
c^{-1}_2; c^{-1}_3\} =  \{(9F - 8)/3; F; 2\}$. 

In general, $c^{-1}_{2(3)}=1,2,\dots f$, where $f$ is the number of
fundamental representations of $SU(2)_L$ $(SU(3)_c)$ contained in the
fundamental representation of the GUT group. For example, $c_2^{-1}=4$ in
$SU(16)$ because the 16 representation of $SU(16)$ contains four $SU(2)_L$
doublets; three for $(u,d)_L$ and one for $(\nu_e,e)_L$. 

The group $[SU(4)]^3\times Z_3$ in Table I is not the vector-like color version
of  the two family Pati-Salam group,  but it is the one family theory
introduced in Ref.~\cite{2so10}. Also, the  group $[SU(6)]^4\times Z_4$ in the
Table is not the 3 family Pati-Salam model,  but a version of such model  (with
3 families) without mirror fermions, introduced in Ref.~\cite{4su6}.

All models in Table I are realistic, except $E_7$\cite{e7} which is a two
family model with the right handed quarks in $SU(2)_L$ doublets.

The values $c_i$ (and Table I) are interesting by 
themselves because they are related to the Kac-Moody 
levels ($\kappa_i$) of String GUTs~\cite{kacmody}. 
Indeed: $c^{-1}_i= \kappa_i, \; i=1,2,3$.  Curiously enough, the values
for $c_1$ are
integer multiple of 1/3 for all the known groups, we do not know why.

\begin{table}
\begin{center}
\begin{tabular}{||r|c|c|c||}
\hline
 Group & $c^{-1}_1$  & $c^{-1}_2$ & $c^{-1}_3$\\
 \hline\hline
 Canonical &5/3   &1  &1 \\
 \hline
$SU(5)\otimes  SU(5)$& 13/3  & 1  & 2 \\
\hline
$ SO(10)\otimes  SO(10)$ & 13/3 & 1 & 2  \\
 \hline
 $[SU(6)]^3\times Z_3$ & 14/3 & 3 & 1 \\
 \hline
 $[SU(6)]^4\times Z_4$ & 19/3 & 3 & 2 \\ 
 \hline
 $E_7$  & 2/3 & 2 &  1 \\
\hline
 $[SU(4)]^3\times Z_3$ & 11/3 & 1 & 1 \\
 \hline
 $[SU(2F)]^4\times Z_4$ & $(9F-8)/3$ & $F$ &  2 \\
 \hline
\end{tabular}
\caption{
%{\quote \small 
{$c_1, c_2$ and $c_3$ 
values for most of the GUT models in the literature. The entry
``canonical" is explained in the main text, and $F=1,2,\dots$ stand for
the number of families in that particular model.}}
\end{center}
\end{table}

\pagebreak

\newpage

{\bf Figure captions:}\\[1em]
Figure 1. Allowed values for $H$ and $N_T'$ for the canonical values
$(c_1,c_2,c_3)= ({3\over 5}, 1, 1)$. Notice that the unification scale $M$ is
independent of the value for $N_T'$.\\[1em]
Figure 2. Allowed region for the parameters $c_1$ and $N_T'$ for models with
$c_2={1\over 3}$, $c_3={1\over 2}$ and $H=2$. The cross represent the case
$c_1={3\over 19}$ and $N_T'=31$ disscussed in the main text.\\[1em]
Figure 3. Allowed values for $H$ and $N_T$ for models with $D$ parity at the
$M_R$ scale, and the canonical values $(c_1,c_2,c_3)= ({3\over 5}, 1, 1)$.
\\[1em]
Figure 4. Allowed region for the parameters $c_1$ and $N_T$ for models with
$c_2=c_3=1$, $H=2$ and $D$ parity above the $M_R$ scale.

\newpage

%%%%%%%%%%%%%%%%%%%%%%
\begin{figure}%[ht]
\centerline{
\epsfxsize=300pt
\epsfbox{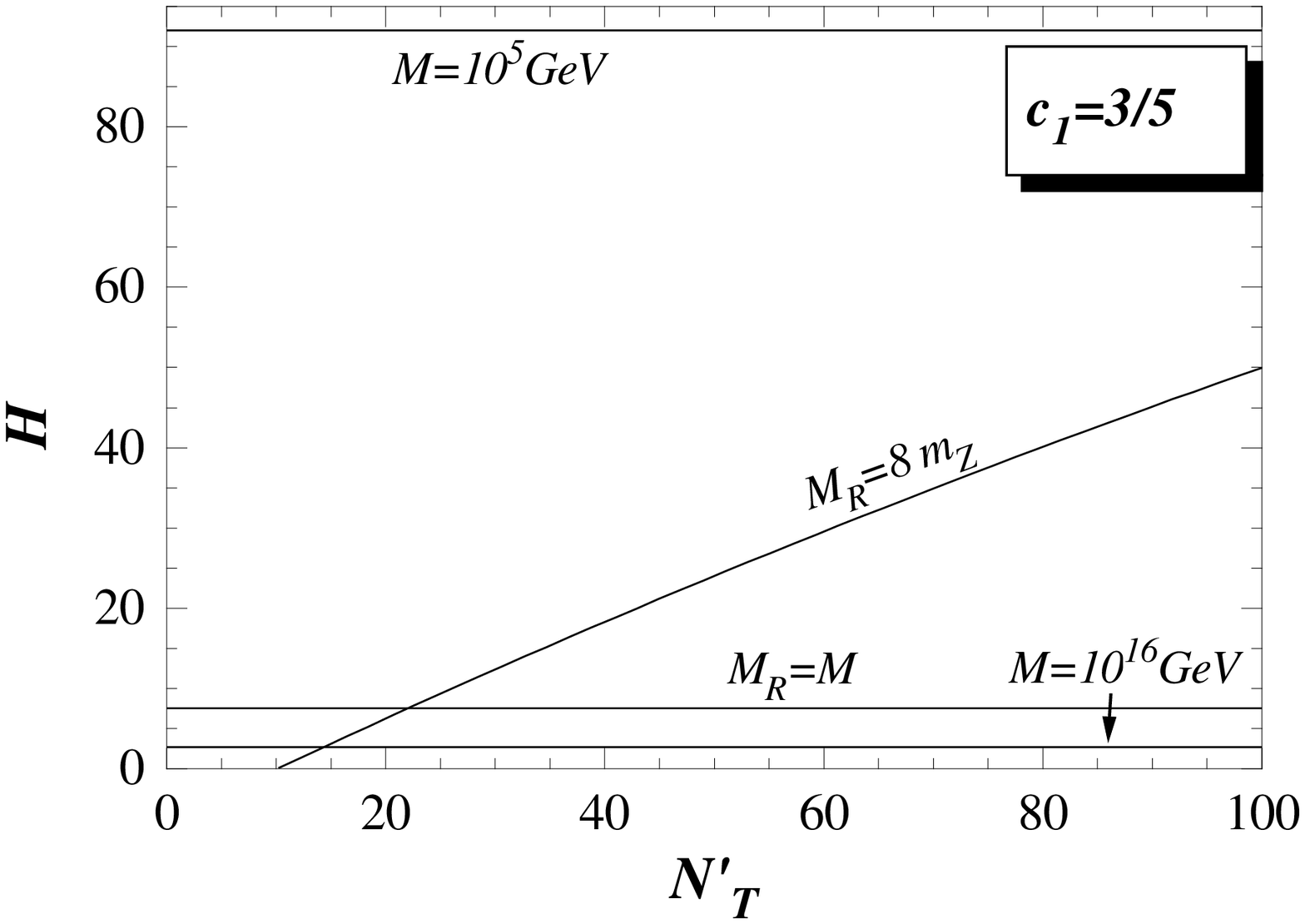}
}
\caption{}
\end{figure}
%%%%%%%%%%%%%%%%%%%%
\vskip1em

%%%%%%%%%%%%%%%%%%%%%%
\begin{figure}%[ht]
\centerline{
\epsfxsize=300pt
\epsfbox{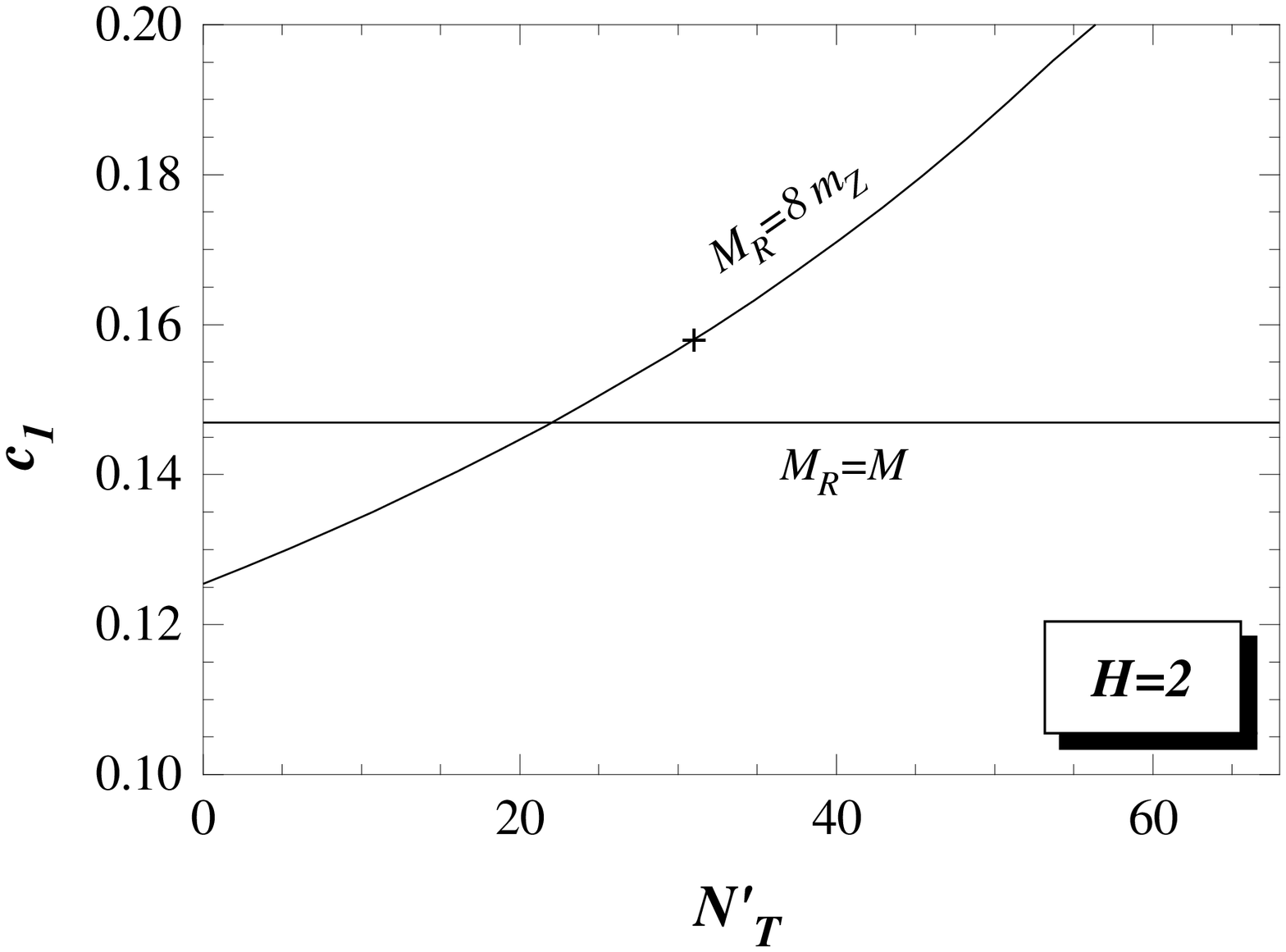}
}
\caption{}
\end{figure}
%%%%%%%%%%%%%%%%%%%%
\vskip1em

%%%%%%%%%%%%%%%%%%%%%%
\begin{figure}%[ht]
\centerline{
\epsfxsize=300pt
\epsfbox{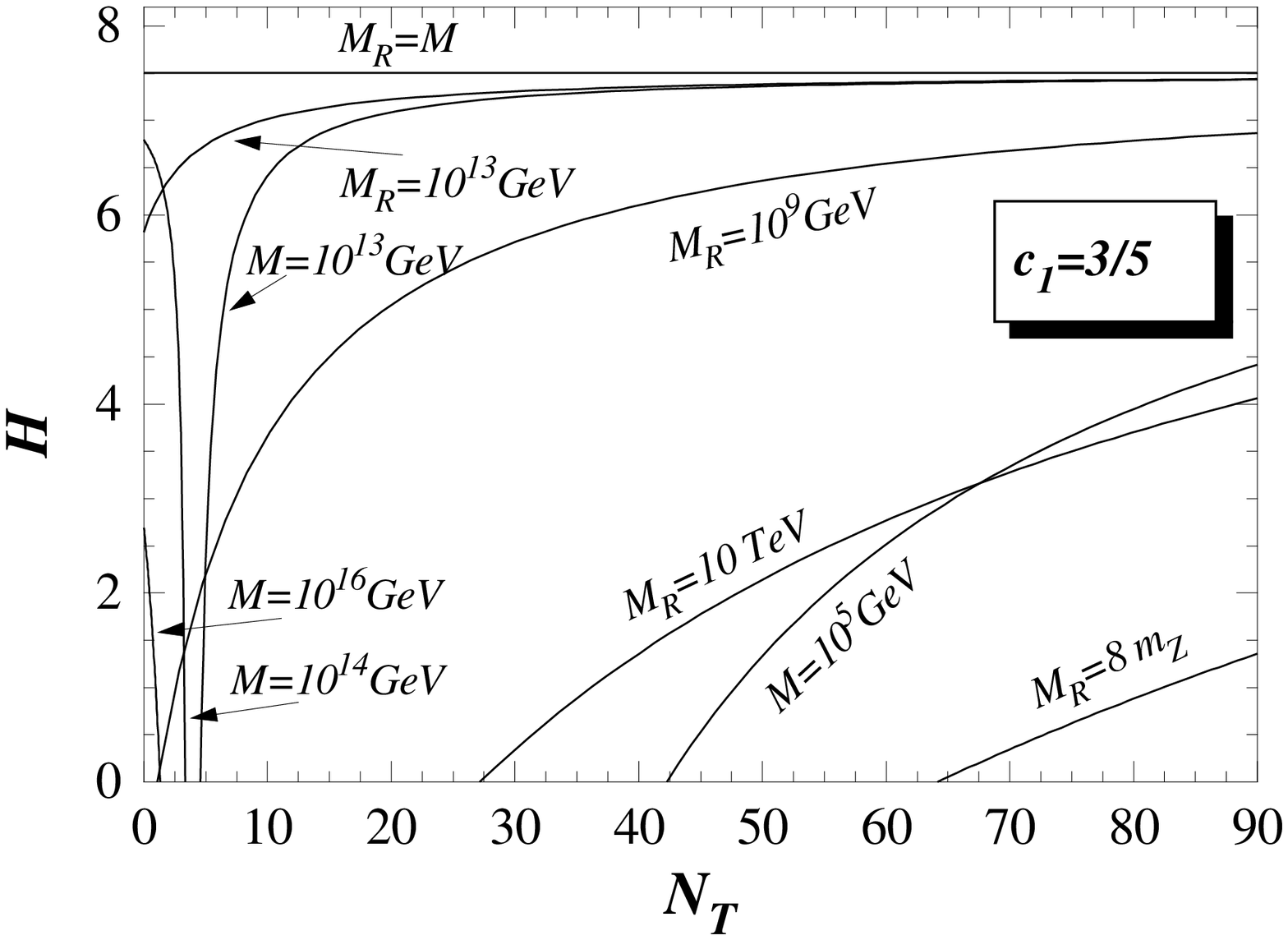}
}
\caption{}
\end{figure}
%%%%%%%%%%%%%%%%%%%%
\vskip1em
%%%%%%%%%%%%%%%%%%%%%%
\begin{figure}%[ht]
\centerline{
\epsfxsize=300pt
\epsfbox{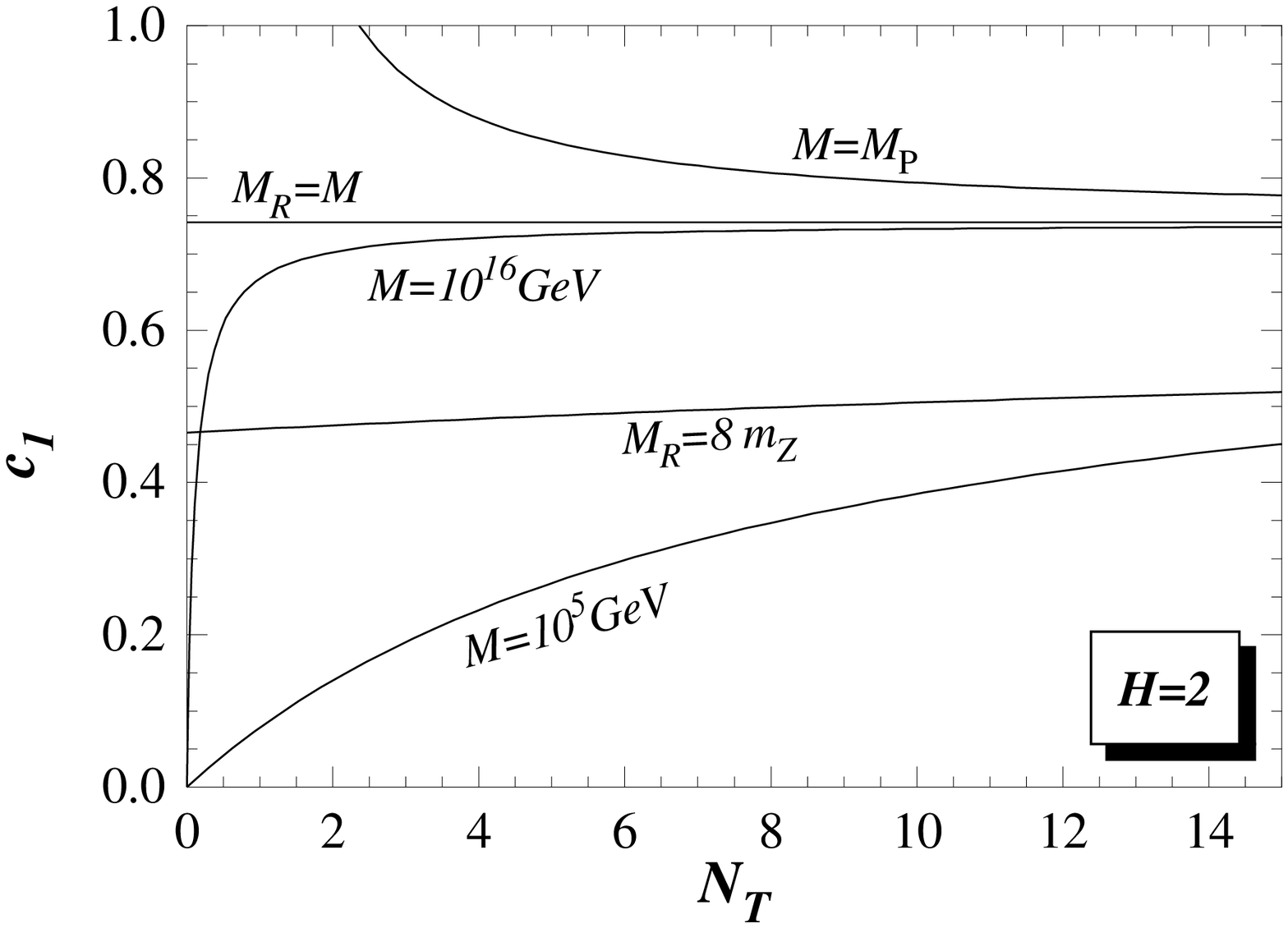}
}
\caption{}
\end{figure}
%%%%%%%%%%%%%%%%%%%

\end{document}